\documentclass{vgtc}
\graphicspath{{figures/}{pictures/}{images/}{./}} 

\usepackage{times}
\usepackage{tabu}
\usepackage{booktabs}
\usepackage{lipsum}
\usepackage{mwe}                     
\usepackage{mathptmx}                  
\usepackage{makecell}
\usepackage{amsmath,amssymb,amsfonts}
\usepackage{algorithmic}
\usepackage{graphicx}
\usepackage{textcomp}
\usepackage{tabularx}
\usepackage{float}
\usepackage{siunitx}

\def\BibTeX{{\rm B\kern-.05em{\sc i\kern-.025em b}\kern-.08em
    T\kern-.1667em\lower.7ex\hbox{E}\kern-.125emX}}

\onlineid{0}
\vgtccategory{Research}
\vgtcinsertpkg

\title{Shaping the Future of VR Hand Interactions: Lessons Learned from Modern Methods}
\author{ByungMin Kim\thanks{e-mail: kbmstar1@korea.ac.kr}\\ %
        \scriptsize Korea University %
\and DongHeun Han\thanks{e-mail: hand32@khu.ac.kr}\\ %
     \scriptsize Kyunghee University %
\and HyeongYeop Kang\thanks{e-mail: siamiz\_hkang@korea.ac.kr (Corresponding Author)}\\ %
     \parbox{1.4in}{\scriptsize \centering Korea University}}
\teaser{
  \centering
  \includegraphics[width=\linewidth]{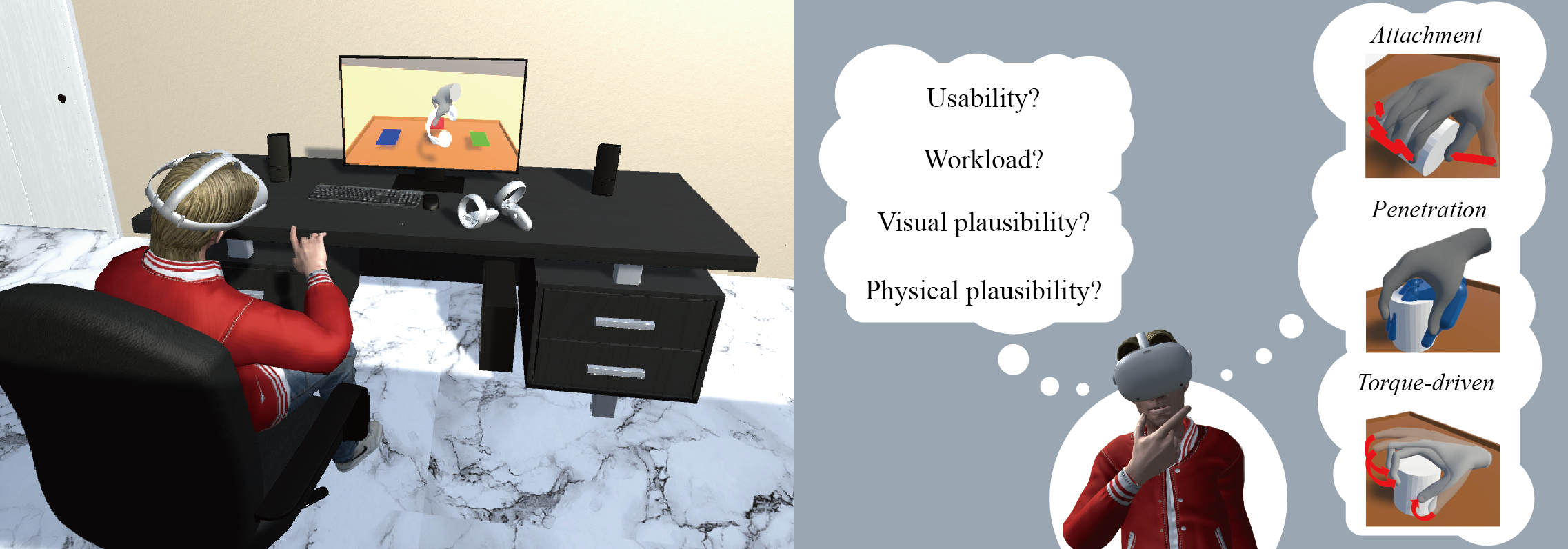}
  \caption{Exploring the strengths and weaknesses of VR hand interaction methods, this study highlights key considerations for commercial applications and proposes improvements for future advancements. Discover how different techniques compare regarding usability, workload, visual plausibility, and physical plausibility.}
  \label{fig:teaser}
}

\abstract{
    In virtual reality (VR), it is widely assumed that increased realism in hand-object interactions enhances user immersion and overall experience. However, recent studies challenge this assumption, suggesting that faithfully replicating real-world physics and visuals is not always necessary for improved usability or immersion. This has led to ambiguity for developers when choosing optimal hand interaction methods for different applications. Currently, there is a lack of comprehensive research to resolve this issue.
    This study aims to fill this gap by evaluating three contemporary VR hand interaction methods—\textit{Attachment}, \textit{Penetration}, and \textit{Torque}—across two distinct task scenarios: simple manipulation tasks and more complex, precision-driven tasks. By examining key technical features, we identify the strengths and limitations of each method and propose development guidelines for future advancements. Our findings reveal that while \textit{Attachment}, with its simplified control mechanisms, is well-suited for commercial applications, \textit{Penetration} and \textit{Torque} show promise for next-generation interactions. The insights gained from our study provide practical guidance for developers and researchers seeking to balance realism, usability, and user satisfaction in VR environments.
}


\begin{document}

\firstsection{Introduction}

\maketitle

In virtual reality (VR), hand interaction methods that enable users to manipulate virtual objects through their avatar's hands represent a highly promising technique. Since hands are our primary means of interacting with the physical world, accurately replicating this functionality in VR facilitates intuitive object manipulation, enhances the sense of realism, and increases user immersion. This approach effectively narrows the gap between real and virtual environments, leading to more natural and engaging user experiences.

In recent years, extensive research has focused on developing VR hand-interaction methods.
The critical aspects that influence the effectiveness and realism of the interaction of these interfaces are the consideration of realistic visual and physical mappings. 
However, accurately mapping real-world hand interactions to virtual environments remains challenging due to hardware limitations. As a result, researchers have adopted approximation methods to balance simplicity and a convincing user experience.

Based on how faithfully these movement and physical properties are mapped in single-hand single-object manipulation, referred to as levels of visual and physical mappings, we identified three types of recent interfaces: Attachment with Controller~\cite{oprea2019visually}, Penetration with Tracking~\cite{quan2020realistic}, and Torque-driven with Controller~\cite{han2023vr} methods. 
The Attachment with Controller method (\textit{Attachment}), a controller-based approach, offers the lowest level of visual and physical mapping. When the user presses a controller button, virtual fingers curl, and upon making contact points with an object, the fingers attach to the object. This method relies on collision detection to simulate the grasping process. However, it disregards physical properties such as force or weight, prioritizing ease of use over physical realism. This approach is well-suited for scenarios where operational stability and simplicity are key requirements.

In contrast, the Penetration with Tracking method (\textit{Penetration}), a tracking-based approach, provides a high level of visual mapping and a moderate level of physical mapping. Inverse kinematics and vision-based techniques are used to accurately map the user’s real hand movements to the virtual hand. Once the virtual hand penetrates the object, forces are calculated based on the degree of penetration, simulating interactions such as squeezing or pressing. This allows for a more realistic interaction by simulating some aspects of physical forces between the hand and the object.

The Torque-driven with Controller method (\textit{Torque}), another controller-based approach, provides a moderate level of visual mapping alongside the highest level of physical mapping. This method generates visually realistic finger movements based on the torque applied to the fingers. The amount of force exerted is determined by the object’s shape, the virtual hand’s position, and the user’s input intensity through the controller. Reinforcement learning and neural network-based models, trained on the authors' custom hand-object interaction dataset, further enhance its capabilities by enabling adaptive control and realistic force application across diverse object geometries. A recent study~\cite{han2023vr} has utilized AI models to determine the appropriate amount of torque applied to finger joints, further enhancing the method’s versatility across various object shapes.

While it is widely believed that greater realism in VR leads to better user experience, recent studies~\cite {prachyabrued2014visual, canales2019virtual} suggest this is not always the case. Faithful visual replication does not necessarily improve user engagement. Moreover, plausibly simulating physical interactions still demand high computational resources, which has prevented recent research from conducting fully comprehensive or comparative analyses of physical plausibility across state-of-the-art VR interaction methods.

These challenges, along with the diverse range of available interfaces, make it difficult for developers to determine which hand-interaction approach is most suitable for specific applications.
Furthermore, it remains unclear whether recent technological advances truly improve user experiences or simply address existing technical constraints, underscoring the need for more extensive research in this area.


In summary, this paper proposes design principles and guides future VR research to enhance user experience in diverse manipulation scenarios. We implemented and evaluated three recent interfaces through human experiments on both simple and high-precision tasks, with thorough analysis providing actionable insights for optimizing VR interactions. The paper’s contributions are as follows:
\begin{itemize}
\item\textbf{Comparative analysis}: 
A comprehensive evaluation of three recent hand-object manipulation methods, clarifying their strengths and weaknesses across different VR scenarios.
\item\textbf{Component-level analysis across different scenarios}: 
Clarifying which technical components offer advantages and disadvantages in two types of VR tasks.
\item\textbf{Practical design guidelines}: 
Practical considerations to optimize user experience, improving both realism and task efficiency in diverse VR interactions.
\item\textbf{Future research directions}: 
Proposing research pathways for each type of interface, identifying which technical characteristics should be preserved and how others can be refined to enhance user experience.
\end{itemize}

\section{Related Works}

\subsection{Comparative Studies of VR Hand-Object Interaction}

Effective hand-object manipulation in VR requires precise tracking of hand and object poses. While handheld controllers were traditionally used, recent advancements in HMDs and vision techniques have enhanced hand-tracking accuracy, making it a viable alternative. 
This shift has prompted a need for comparative studies on the usability, satisfaction, and immersion of controller-based versus tracking-based methods, emphasizing the importance of understanding their respective strengths and limitations in VR applications.

Previous comparative studies have focused on the basic functionalities. Laukka et al.~\cite{laukka2021comparing} assessed presence, satisfaction, and weight sensation when lifting real and virtual objects using hand tracking and controllers. Khundam et al.~\cite{khundam2021comparative} explored usability in an intubation training task by comparing interactions with surgical tools using both input methods. Kapsoritakis et al.~\cite{kapsoritakis2022comparative} investigated user satisfaction through UI interactions, including button presses and slider adjustments. Similarly, Kim~\cite{kim2022comparative} examined browsing and object manipulation tasks to evaluate usability, workload, and satisfaction with controllers and hand tracking. Kangas et al.~\cite{kangas2022trade} demonstrated that a trade-off exists between task accuracy, task completion time, and naturalness using a comparative study of controllers and hand tracking.

These studies, however, primarily relied on the Controller-Tracking binary classification. We argue that such a binary scheme fails to capture the nuanced ways in which physical and visual properties influence interactions. For instance, two systems classified under “controller-based” might differ considerably in how they simulate finger movement and force exerted by the virtual hand, yet these nuances are masked by a binary labeling scheme. To overcome this limitation and enable a more comprehensive comparison, we choose three recent VR hand interaction methods~\cite{oprea2019visually, quan2020realistic, han2023vr}, each employing distinct design choices in visual and physical plausibility, which influence naturalness and the overall user experience in VR environments.



\subsection{Hand Movement Mapping - Visual Plausibility }
In traditional hand-object manipulation, pre-generated grasping animations tailored for each object were used to satisfy visual plausibility~\cite{li2007data, hamer2011data}. This method had the advantage of simplifying the complex interactions between the hand's degrees of freedom and the object's geometry. However, the limitation was that it could only represent one or a predefined number of grasping animations, and it was not capable of generating multiple grasping poses in real-time based on the region of the object being grasped. Crucially, this method also had the drawback of being unable to represent visually plausible grasping for unseen objects.

With the development of HMDs, real-time tracking of the user's hands became possible, and various methods were attempted to create visually plausible hand-object manipulation for unseen objects using controllers and hand tracking. For instance, Shi et al.~\cite{shi2022grasping} and Oprea et al.~\cite{oprea2019visually} proposed a visually plausible grasp synthesis method using the blocking effect caused by collisions between the hand model, generated by finger-bending animations via a controller, and the object's colliders. This approach efficiently met the requirements for visual plausibility but had the limitation of not allowing for diverse grip styles, as all fingers were bent simultaneously with a single controller button.

Meanwhile, methods were also studied to allow the user to control the movement of their fingers in real-time through hand tracking while interacting with virtual objects in a visually plausible manner. However, even with hand tracking, the user's real hands cannot physically contact virtual objects, inevitably leading to interpenetration. Therefore, most studies forced the rendered hand to appear as if it were exactly touching the contact point to prevent interpenetration. For example, H\"{o}ll et al.~\cite{holl2018efficient} presented a method where the hand’s pose freezes once it touches the object, creating the appearance of the hand exactly contacting the point. Additionally, Quan et al.~\cite{quan2020realistic} utilized inverse kinematics to adjust each joint's angles according to contact point changes, creating visually plausible motion.

\subsection{Physical Property Mapping - Physical Plausibility }

In the real world, we can effortlessly perform physical interactions such as grasping or moving objects without much thought. This is because we automatically perceive the distance and force exerted on objects based on visual distance information and tactile pressure information. However, it is challenging to spatially perceive the shapes of objects and hands in virtual reality due to the limitations of the field of view, and because the objects are virtual, methods for perceiving weight or contact are very limited and costly. Therefore, mapping physical properties in hand-object manipulation is both a crucial and challenging topic. 

Many studies have attempted to analyze the forces exerted by the hand on the surface of the object. Holz et al.~\cite{holz2008multi} and Moehring et al.~\cite{moehring2010enabling} utilized Coulomb friction cones and grasp pairs acting on the contact points from both directions relative to the object's center of mass to produce stable grasping. Li et al.~\cite{li2011simulation} also attempted to model the friction occurring between the fingers and the object using Coulomb friction cones. However, these methods failed to properly handle the issue of interpenetration with virtual objects and did not represent interactions between objects.

To address this, Jacobs et al.~\cite{jacobs2012generalized} proposed a method using the "God-hand object approach," which calculates the force applied to the object by the difference between the position of the god-hand object and the tracked hand when interpenetration occurs between the hand and the object. Nasim et al.~\cite{nasim2018physics}, H\"{o}ll et al.~\cite{holl2018efficient} and Quan et al.~\cite{quan2020realistic} extended this by using Coulomb friction models. They calculated the force applied at each contact point as either static or kinetic friction using Coulomb friction cones, producing more accurate physical interactions. However, these methods made it difficult for users to arbitrarily control the amount of penetration, making fine-tuning the force magnitude challenging. Furthermore, they did not consider the equilibrium of the object during the application of force, which decreased the stability of the grasp.

Meanwhile, with the advancement of reinforcement learning and imitation learning, attempts have also been made to learn the physical properties of hand-object manipulation motions from real users. Rajeswaran et al.~\cite{rajeswaran2017learning} combined reinforcement learning and imitation learning to solve complex hand-object manipulation. OpenAI~\cite{andrychowicz2020learning} successfully created a physical in-hand manipulation system by applying deep reinforcement learning to a physical robot hand, using physics parameter randomization during training. Han et al.~\cite{han2023vr} proposed physically plausible hand-object grasping synthesis by approximating and mapping forces based on the bending of the hand and the movement of the object in hand-object manipulation motion data.

\section{Study Design}

In this section, we present the overall design of the study. The study begins with implementing state-of-the-art single-hand single-object interaction methods for testing, including the ``Attachment with Controller method", ``Penetration with Tracking", and ``Torque-driven with Controller method". Following these, two comparative experiments are conducted to investigate our research questions. 

In the first experiment (\textbf{E1}), the implemented methods are evaluated with the ``object relocation" task, which involves the fundamental VR interactions of grasping, moving, rotating, and releasing objects. These basic interactions are frequently utilized in various VR applications.
In the second experiment (\textbf{E2}), the methods are tested with a more complex ``tower-building" task that requires higher cognitive and physical effort to avoid tower collapse and effectively stack the next object.
For both experiments, we utilized complex objects from the GRAB dataset~\cite{taheri2020grab}. Since these objects lacked colliders, we implemented mesh colliders and divided them into sections to ensure accurate collision detection and handling.


\subsection{Implementing Test Hand Interaction Methods}

\begin{figure*}
    \centering
    \includegraphics[width=1\linewidth]{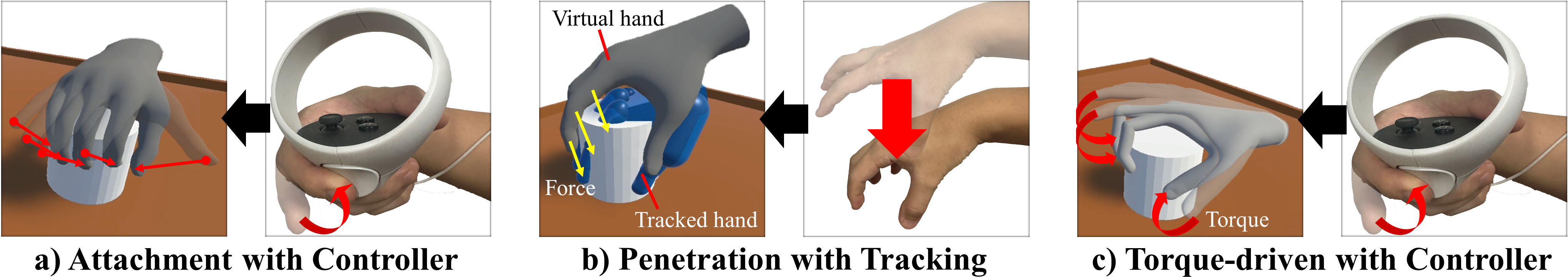}
    \caption{The illustration of three contemporary VR hand interaction methods: (a) Attachment with Controller, (b) Penetration with Tracking, and (c) Torque-driven with Controller.}
    \label{fig:Method_overview}
\end{figure*}

\subsubsection{Attachment with Controller Method} 
Faithfully implementing the method described by Oprea et al.~\cite{oprea2019visually}, \textit{Attachment} enables users to grasp virtual objects by pressing a button on the controller. Attachment enables users to grasp virtual objects by pressing a button on the controller. When the button is pressed, the nearest object overlapping with an invisible spherical trigger collider on the virtual palm is selected. Subsequently, each finger bends progressively toward the object's surface, with the joints flexing sequentially based on collision feedback and predefined angular constraints. The predefined angular constraint ensures that the hand joints move within realistic angles, typically up to 90 degrees, to maintain a natural appearance. A stable grasp is achieved when the system detects sufficient contact points, specifically involving either the thumb or palm along with the index finger or the middle finger. This ensures that the grasp is secure enough to prevent unintentional detachment. Once stable grasp is achieved, the object is anchored to the hand using a fixed joint, remaining in place until the button is released. \autoref{fig:Method_overview}(a) illustrates how \textit{Attachment} works.

\subsubsection{Penetration with Tracking Method}
Faithfully implementing the method described by Quan et al.~\cite{quan2020realistic}, \textit{Penetration} allows users to control the strength of their grasp when interacting with a virtual object while closely mimicking real hand movements. Utilizing vision-based tracking techniques, the virtual hand replicates the real hand's movements until contact is made with a virtual object. Once contact occurs, the difference between the tracked hand's position and the virtual hand's contact points generates forces applied to the object. These forces are refined using inverse kinematics and Coulomb friction models, determining whether the interaction results in a stable (static) grip or a slipping (kinetic) contact. When slippage is detected, the method updates the contact points and recalculates the hand’s pose accordingly.

To achieve a balanced distribution of forces and to represent the deformable nature of the hand, the force is applied to a contact area defined by six points surrounding the primary contact point, rather than being directly assigned to each finger. The normal vector of this contact area aligns with the normal of the primary contact point.
However, this method can sometimes suffer from an imbalance in force distribution, where the four fingers tend to exert more force than the thumb, leading to unintended movements and increased penetration into the virtual object. To mitigate this issue, high drag and angular drag values are applied to minimize oscillatory movements, ensuring stability. These values are also used for consistency in other methods. \autoref{fig:Method_overview}(b) illustrates how \textit{Penetration} works.

\subsubsection{Torque-driven with Controller Method} 
Building upon the pre-trained model developed by Han et al.~\cite{han2023vr}, \textit{Torque} generates realistic finger joint movements to grasp virtual objects while simultaneously producing torque that reflects the intensity with which users press the trigger button. This approach utilizes data-driven deep reinforcement learning to achieve its objectives. Specifically, it employs a pre-trained VR-HandNet model to compute the necessary torque for realistic finger animations and the torque applied to the virtual object. The remaining calculations are then handled by a Proportional-Derivative (PD) controller within the test virtual environment.  \autoref{fig:Method_overview}(c) illustrates how \textit{Torque} works.

\subsubsection{Test Object Implementation} 
Primitive objects such as cubes or tetrahedrons are widely used in VR experiments since they have simple shapes that are likely to yield high grasping performance. However, as these shapes are not commonly encountered in real-world scenarios, their realism may be limited. 
To overcome this, we utilized 50 objects from the well-known full-body 3D object grasping motion dataset, GRAB \cite{taheri2020grab}.

However, the objects in the GRAB dataset are provided as meshes without colliders, making them unsuitable for collision-based interactions. To resolve this, we manually created mesh colliders that closely aligned with the original mesh shapes, enabling accurate collision responses. Specifically, we divided the meshes into multiple parts and assigned mesh colliders to each section, ensuring they collectively functioned as a unified collider. This approach allowed the objects to interact realistically during interactions while maintaining their original shapes.

\subsection{Research Goals}
While it is commonly believed that the accuracy of physical interactions in hand-manipulation tasks enhances user immersion, several researchers ~\cite{canales2019virtual, prachyabrued2014visual, slater2022consciousness, zhou2024grasping} argue that in VR, faithfully replicating the real world with visual, physical properties are not always necessary for achieving high level of immersion. Instead, they argue that creating virtually plausible environments that effectively engage the user's consciousness is more critical, even if these environments do not perfectly mirror reality. 

Despite the ongoing debate about whether faithfully replicating the visual animations and physical plausibility is essential for achieving higher immersion in VR hand interaction methods, and to what extent such replication is necessary, no study has yet provided a clear answer. This gap in the research persists because VR hand interaction methods are still in their developmental stages, with most studies focusing primarily on overcoming known technical challenges.

To resolve such ambiguity, we examine user experiences with three recent VR hand interaction methods in different interaction scenarios. Here, ``user experience" encompasses multiple dimensions, including usability, task efficiency, workload, and user preference. 
Our study seeks to address two key research goals:
\begin{itemize}
\item\textbf{Goal 1}: To identify the technical features that contribute to the strengths and weaknesses of VR hand interaction methods when interacting with diverse objects and scenarios.
\item\textbf{Goal 2}: To propose development guidelines for future advancements, specifying which technical characteristics should be retained and how others can be refined for improved user experiences.
\end{itemize}

By examining \textbf{Goal 1}, we aim to provide component-level insights into how specific features of each method influence its effectiveness and limitations in VR hand interactions. This analysis will offer an objective assessment of the commercial viability of current VR hand interaction methods, establishing a benchmark for their readiness for real-world implementation. Additionally, the findings will offer practical guidance for content developers in selecting the most suitable methods for various VR applications.

In exploring \textbf{Goal 2}, we aim to identify how each method can be refined further, offering direction for future research that goes beyond solving known technical issues to enhance the overall user experience.

\begin{table}[h]
\caption{Naturalness questionnaire.}
\label{tab:NQ_Questions}
\setlength{\tabcolsep}{3pt}
\begin{tabularx}{\linewidth}{>{\hsize=.43\hsize}X|>{\hsize=.57\hsize}X}
\Xhline{2\arrayrulewidth}
Category & Question \\ \Xhline{2\arrayrulewidth}
Hand Naturalness \newline (HN)& The hand I manipulated felt like my own.                                                  \\ \hline
Object Naturalness \newline  (ON)& The objects moved by my hand felt like real objects.                                      \\ \hline
Manipulation Naturalness \newline  (MN) & While performing the task, it felt like I was actually manipulating the objects.          \\ \hline
Visual plausibility \newline  (VP)& The hand used to grasp objects behaved similarly to how I would grasp objects in reality. \\ \hline
Physical plausibility \newline  (PP)& It felt like the force I applied was accurately transmitted to the virtual hand.          \\ \Xhline{2\arrayrulewidth}
\end{tabularx}
\end{table}

\section{Experiment 1: Object Relocation Task}
The relocation task used in E1 requires participants to grasp an object, rotate it to examine information engraved on its surface, and then move the object to a designated location according to the retrieved information. This task requires fundamental hand interactions, including grasping, moving, rotating, and releasing objects. However, it does not demand high levels of physical precision or fine motor control from the user.

\subsection{Participants and Apparatus}
The study involved 24 participants, comprising 19 males and 5 females. Of these, 19 participants had prior experience with VR, while 5 had no previous exposure. The $\mu$ and $\sigma$ of age were 26.083 $\pm$ 0.584. The experiment was conducted using an Oculus Quest 2 headset, paired with controllers, and operated on a computer equipped with an RTX 3070 graphics card and an AMD Ryzen 7 3800XT processor.

\subsection{Settings and Procedure}
E1 was conducted with two experimenters. Upon arrival, each participant was asked to complete a consent form and a demographics questionnaire. Participants were then given a training session, lasting up to 15 minutes, to familiarize themselves with the experimental setup. This session included a 5-minute overview of the experiment, followed by up to 10 minutes of free practice.

After the training, the main experiment began. Participants were asked to fill out a Simulator Sickness Questionnaire (SSQ)~\cite{kennedy1993simulator}. Then they completed three trials, with each trial testing a different hand interaction method. The order in which the interaction methods were tested was counterbalanced across participants to minimize bias.

At the beginning of each trial, a single object and three colored boxes labeled R, G, and B were generated. The object appeared at a fixed location, whereas the colored boxes were randomly placed within nine designated areas: inside, middle, or outside (relative to the participant's position); and left, middle, or right (relative to the participant's body orientation). These areas were configured to avoid overlap among the boxes.

Each object's shape was randomly selected from 50 possible shapes provided by GRAB dataset~\cite{taheri2020grab}, and its orientation was randomly determined from 512 possible directions, defined by three rotational axes with 45-degree increments.
The size of these shapes ranged from approximately 6-18cm in diameter, ensuring that objects were neither too small to be easily overlooked nor too large to handle.

Participants were required to grasp the object, rotate it to identify the color-coded letter on one of its surfaces, and then move the object to the corresponding colored box. 
They had up to 20 seconds to complete each classification task. Whenever a participant successfully classified an object or failed to do so within 20 seconds, the system generated a new object at the fixed position, and the three colored boxes were relocated to random positions within the nine designated areas. Each trial lasted a maximum of 180 seconds. During these, participants could classify as many objects as possible in succession.

After completing each trial, participants were asked to complete the Simulation-Task Load Index (SIM-TLX)~\cite{harris2020SIMTLX}, System Usability Scale (SUS)~\cite{brooke1996sus}, Igroup Presence Questionnaire (IPQ)~\cite{schubert2003sense}, and the NQ, which is detailed in~\autoref{tab:NQ_Questions}. Because earlier studies typically measured naturalness using a single questionnaire item~\cite{holl2018efficient, shi2022grasping}—which does not distinguish the specific elements that shape perceptions of naturalness—we developed a customized NQ comprising five dimensions: hand, object, interaction, visual plausibility, and physical plausibility.
At the end of all three trials, participants were asked to complete the SSQ again. An open-ended interview was then conducted to gain deeper insights into their experiences.

\begin{figure}
    \centering
    \includegraphics[width=1\linewidth]{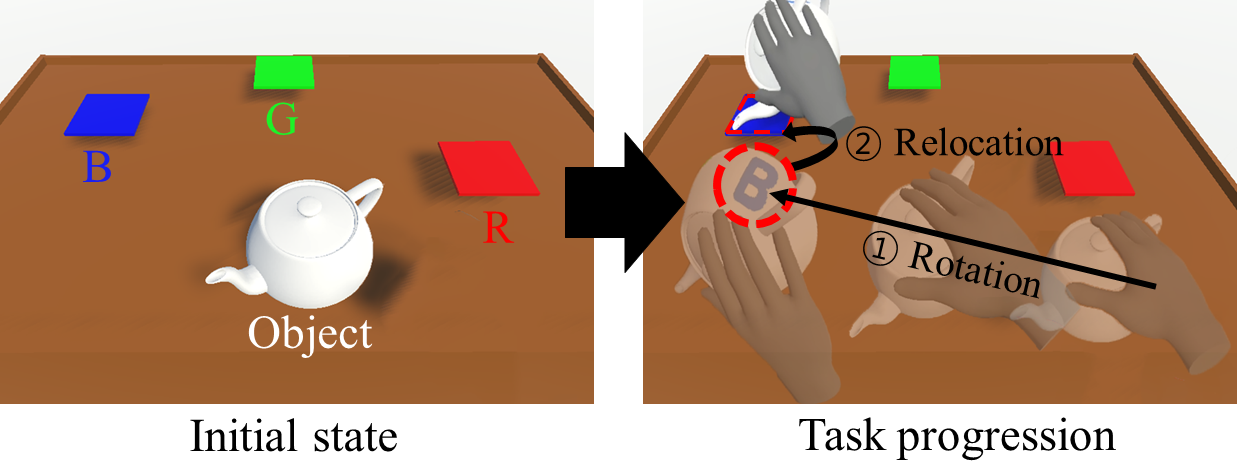}
    \caption{Illustration of the relocation task: Participants rotate the object to identify a color-coded letter (R, G, or B) and then move the object to the corresponding colored box.}
    \label{fig:Overview_object_relocation_task}
\end{figure}

\subsection{Result}

\begin{table}[h]
    \caption{The recorded performance of participants across the three different test methods in E1.}
    \label{tab:Object_relocation_task}
    \centering
    \resizebox{\linewidth}{!}{%
    \begin{tabular}{lSSS}
        \Xhline{2\arrayrulewidth}
        {} & {\textit{Attachment}} & {\textit{Penetration}} & {\textit{Torque}} \\
        \Xhline{2\arrayrulewidth}
        Task completion time (s) & 4.735 & 8.143 & 7.616 \\ 
        \hline 
        \# of object drops & 1.814 & 3.247 & 1.819 \\ 
        \hline 
        \# of relocation successes & 38.958 & 14.125 & 17.727 \\ 
         \hline 
        \# of relocation fails & 2.833 & 3.792 & 3.958 \\ 
        \Xhline{2\arrayrulewidth}
    \end{tabular}
    }
\end{table}

\begin{table}[h]
    \caption{Mean and standard of SIM-TLX in E1.}
    \label{tab:Object_relocation_task_TLX}
    \centering
    \resizebox{\linewidth}{!}{%
    \begin{tabular}{l||SSS}
        \Xhline{2\arrayrulewidth}
        {Dimension} & {\textit{Attachment}} & {\textit{Penetration}} & {\textit{Torque}} \\
        \Xhline{2\arrayrulewidth}
        PD & {31.12 $\pm$ 7.02} & {97.88 $\pm$ 16.03} & {77.63 $\pm$ 11.86} \\ 
        \hline 
        MD & {18.34 $\pm$ 6.21} & {78.00 $\pm$ 14.11} & {48.00 $\pm$ \phantom{0}9.69} \\ 
        \hline 
        FR & {\phantom{0}7.67 $\pm$ 4.79} & {34.33 $\pm$ 12.58} & {23.67 $\pm$ \phantom{0}8.87} \\ 
        \hline 
        DI & {\phantom{0}2.08 $\pm$ 2.59} & {\phantom{0}4.50 $\pm$ \phantom{0}5.53} & {\phantom{0}2.83 $\pm$ \phantom{0}2.91} \\ 
        \hline 
        CO & {14.50 $\pm$ 6.50} & {65.75 $\pm$ 15.92} & {47.25 $\pm$ 13.25} \\ 
        \hline 
        TC & {\phantom{0}3.83 $\pm$ 3.20} & {\phantom{0}9.33 $\pm$ \phantom{0}6.18} & {11.50 $\pm$ \phantom{0}7.12} \\ 
        \hline 
        TD & {\phantom{0}9.50 $\pm$ 7.64} & {26.25 $\pm$ 11.34} & {13.88 $\pm$ \phantom{0}8.57} \\ 
        \hline 
        PS & {21.88 $\pm$ 6.85} & {46.25 $\pm$ 13.84} & {35.63 $\pm$ 12.57} \\ 
        \hline 
        SS & {15.75 $\pm$ 5.69} & {63.58 $\pm$ 15.91} & {38.50 $\pm$ \phantom{0}8.65} \\ 
        \Xhline{2\arrayrulewidth}
    \end{tabular}
    }
\end{table}

\begin{figure}
    \centering
    \includegraphics[width=1\linewidth]{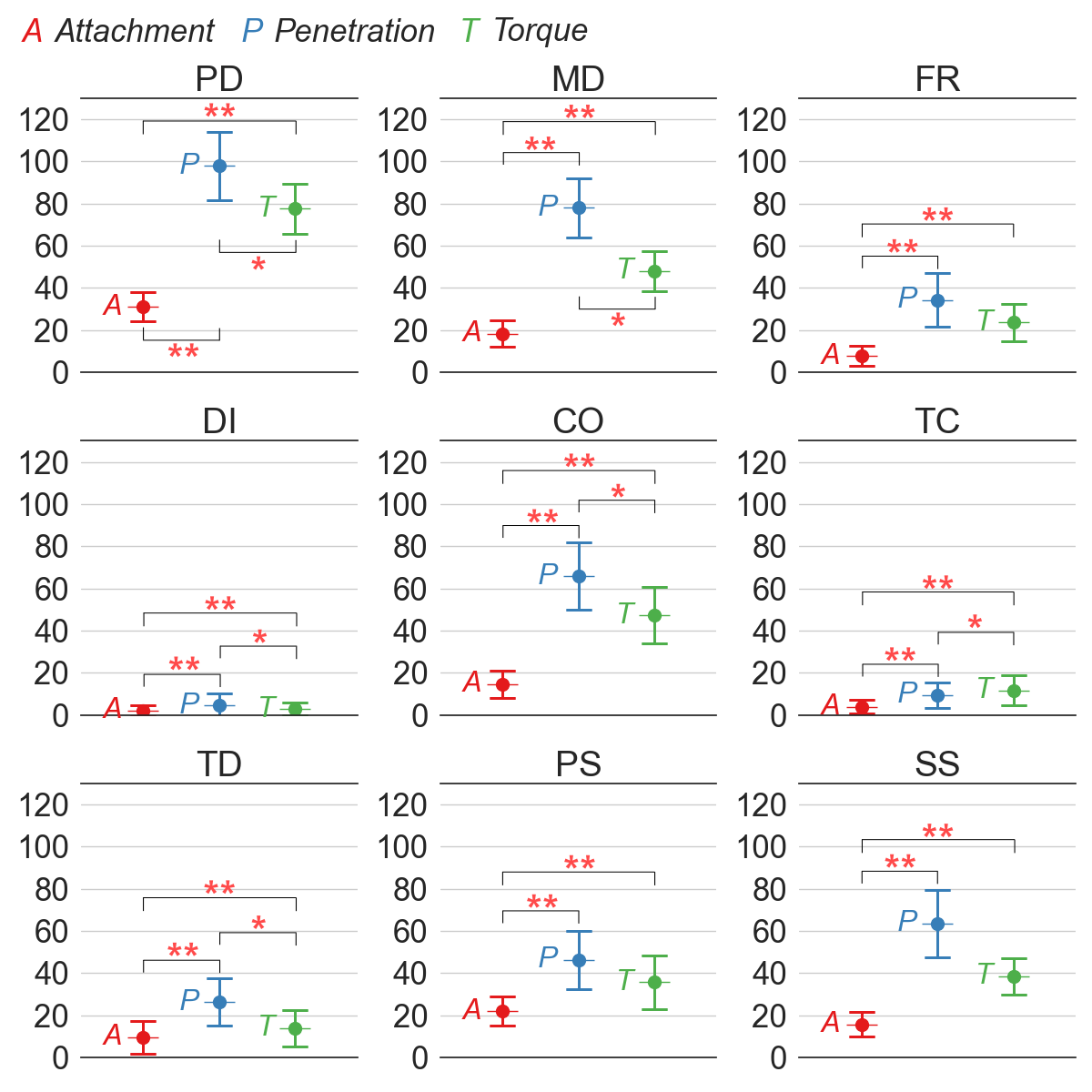}
    \caption{SIM-TLX analysis results for object relocation task. The graph plots the mean and the standard deviation. Square brackets between groups within the same item indicate the results of the Wilcoxon signed-rank test ($*$ : p $<$ 0.05, $**$ : p $<$ 0.01). }
    \label{fig:SIM-TLX_ObjectRelocationTask}
\end{figure}

\begin{figure*}[h]
    \centering
    \includegraphics[width=\linewidth]{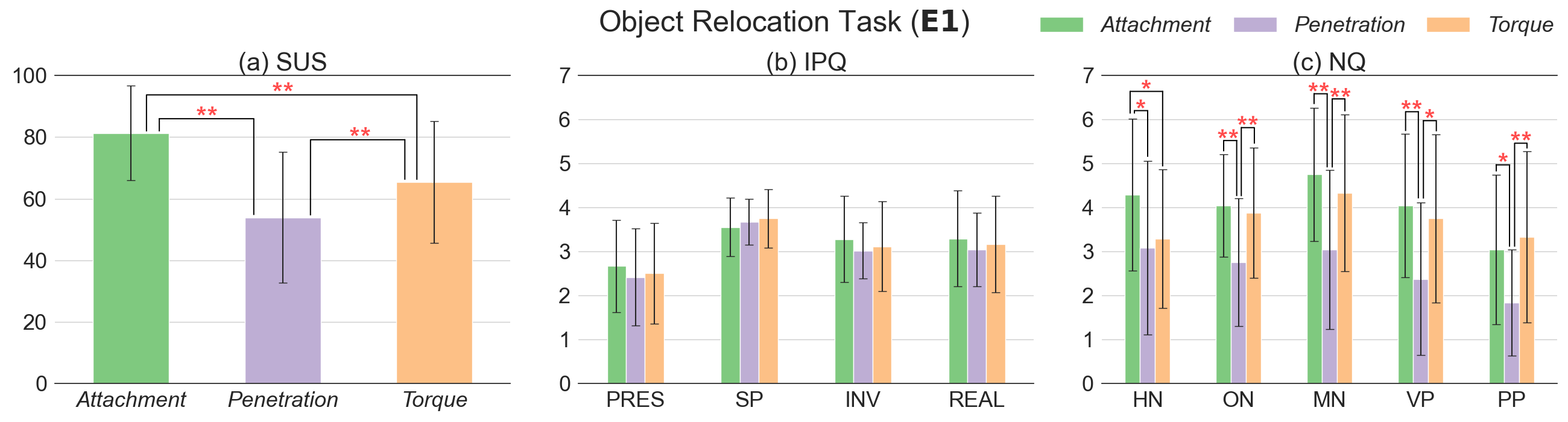}
    \caption{Result of SUS, IPQ, and NQ. The graph plots the mean and the standard deviation. Square brackets between groups within the same item indicate the results of the Wilcoxon signed-rank test ($*$: p $<$ 0.05, $**$ : p $<$ 0.01). }
    \label{fig:Subjective_ObjectRelocationTask}
    \vspace{-1mm}
\end{figure*}

\subsubsection{Objective Evaluation} 
\autoref{tab:Object_relocation_task} summarizes the participants' performance recorded across three different test methods in the object relocation task scenario. The performance is evaluated in terms of task completion time, defined as the average time participants took to complete a single relocation task, the average number of object drops, the average number of relocation successes, and the number of failed relocation attempts recorded during the experiment.  

Overall, \textit{Attachment} demonstrates superior performance compared to \textit{Torque} and \textit{Penetration}. 
This advantage is likely due to its simplicity, as it focuses only on the relative positions of the hand and the object during grasping, without requiring users to account for the force they need to apply. Additionally, the use of a controller enhances operational precision, thereby minimizing the likelihood of errors. This precision benefit also contributes to the \textit{Torque}'s improved performance over \textit{Penetration} in terms of fewer object drops and a higher number of successful interactions.


\subsubsection{Subjective Evaluation}
To analyze the collected responses, we first assessed normality using the Shapiro-Wilk test. Since some responses did not follow a normal distribution, we applied the Friedman test to detect statistically significant differences. For post-hoc analysis, we conducted Wilcoxon signed-rank tests.

\autoref{fig:SIM-TLX_ObjectRelocationTask} and \autoref{tab:Object_relocation_task_TLX} are the SIM-TLX analysis results that demonstrate the user's workload through nine dimensions: Physical demand (PD), Mental demand (MD), Frustration (FR), Distraction (DI), Task control (CO), Task complexity (TC), Temporal demand (TD), Perceptual strain (PS), and Situational stress (SS). The Friedman test results indicated significant differences across all evaluated dimensions: PD, MD, TD, SS, TC, and FR ($p < 0.01$), as well as DI and PS ($p < 0.05$).  Notably, \textit{Penetration} consistently scored higher than the other two methods except for the TC dimension.

\autoref{fig:Subjective_ObjectRelocationTask}(a) presents the SUS analysis. It revealed a ranking of usability with \textit{Attachment} demonstrating the highest usability ($\mu$ = $81.354$, $\sigma$ = $\pm$15.304), followed by \textit{Torque} ($\mu$ = $65.417$, $\sigma$ = $\pm$19.791), and \textit{Penetration} ($\mu$ = $53.854$, $\sigma$ = $\pm$21.264).
Statistical comparisons revealed that \textit{Attachment} had significantly higher usability scores than both \textit{Penetration} and \textit{Torque}. Additionally, \textit{Torque} showed higher usability scores than \textit{Penetration} ($\chi^2(2) = 23.383$, $p < 0.01$). As previously discussed, the simplicity of focusing solely on grasping based on the relative positions of the hand and the object, along with the use of a controller, appears to have positively influenced usability.

\autoref{fig:Subjective_ObjectRelocationTask}(b) is the IPQ analysis results that demonstrate users' sense of presence through four dimensions: overall presence (PRES), spatial presence (SP), involvement (INV), and perceived realism (REAL). Friedman test revealed that none of the dimensions showed statistical significance.

\autoref{fig:Subjective_ObjectRelocationTask}(c) presents an analysis of the results, reporting naturalness across five dimensions: Hand Naturalness (HN), Object Naturalness (ON), Hand-Object Manipulation Naturalness (MN), Visual Plausibility (VP), and Physical Plausibility (PP).
For the HN dimension, \textit{Attachment} ($\mu$ = 4.292, $\sigma$ = $\pm$1.732) scored significantly higher than \textit{Penetration} ($\mu$ = 3.083, $\sigma$ = $\pm$1.976) and \textit{Torque} ($\mu$ = 3.292, $\sigma$ = $\pm$1.574) ($\chi^2(2) = 9.976$, $p < 0.01$) 
For the ON dimension, \textit{Attachment} ($\mu$ = 4.042, $\sigma$ = $\pm$1.160) scored higher than \textit{Penetration} ($\mu$ = 2.750, $\sigma$ = $\pm$1.452) and \textit{Torque} ($\mu$ = 3.875, $\sigma$ = $\pm$1.484)($\chi^2(2) = 13.730$, $p < 0.01$). 
For the VP dimension, \textit{Attachment} ($\mu$ = 4.042, $\sigma$ = $\pm$1.628) scored higher than \textit{Penetration} ($\mu$ = 2.375, $\sigma$ = $\pm$1.740) and \textit{Torque} ($\mu$ = 3.750, $\sigma$ = $\pm$1.917) ($\chi^2(2) = 13.705$, $p < 0.01$). 
Lastly, for the PP dimension, \textit{Torque} ($\mu$ = 3.333, $\sigma$ = $\pm$1.949) scored higher than \textit{Penetration} ($\mu$ = 1.833, $\sigma$ = $\pm$1.830) and \textit{Attachment} ($\mu$ = 3.042, $\sigma$ = $\pm$1.706) ($\chi^2(2) = 12.795$, $p < 0.01$). The MN dimension did not show statistical significance.

There was no statistically significant change in SSQ scores before and after the experiment.




\section{E2: Tower-Building Task}
The tower-building task used in E2 requires participants to stack virtual objects with a high degree of precision. This task requires users not only to perform basic hand interactions but also to carefully adjust the placement of each object to avoid tower collapse due to the force exerted by the hand or object imbalance. Consequently, participants must consider multiple factors,  including the contact points between hand and object, the timing of release, the size and shape of the objects, and the order in which they are stacked. After a 10-minute break, the participants from E1 proceeded to participate in E2.

\subsection{Settings and Procedure}
E2 was conducted with two experimenters using a similar protocol to E1. First, participants completed a 15-minute training session (5-minute overview plus up to 10 minutes of free practice) to familiarize themselves with the E2. After the training, participants filled out the SSQ, establishing a baseline for simulator sickness measures. Participants performed three separate trials, each testing a different hand interaction method. The order of the interaction methods was counterbalanced across participants to minimize bias.

At the beginning of each trial, two objects were randomly generated within a designated area on a virtual table. 
The objects were limited to cylinders and cubes, each with simple planar meshes to facilitate grasping and stacking. Three sizes (small, medium, large) were used, where medium-sized objects ranged from approximately 10-12cm in diameter, and small and large objects were defined by multiplying the medium dimension by 0.5 and 1.5, respectively. This variability required participants to adapt their grasping strategy and consider object size when stacking. As in Experiment 1, each object’s orientation was randomly chosen from 512 possible directions.

Participants first stacked one object atop another. Once the second object was placed, they were required to remove their virtual hand from the object and maintain a stable tower for five seconds (i.e., the objects should not tip over or fall). If the tower successfully remained stable for the designated five seconds, the scenario was reset: the existing tower was removed, and three objects of varying shapes and sizes were randomly repositioned within a designated area on the virtual table. With each successful completion of the stacking task, an additional object was introduced, gradually increasing the challenge by requiring greater precision and strategic planning to maintain stability. 

Each trial lasted a maximum of 180 seconds including both the stacking actions and the 5-second stability checks. The 5-second stability check was not an additional enforced wait before moving on; rather, it was the time during which the tower’s stability was verified. If the tower fell during or after these five seconds, participants had to restart the tower with the same set of objects (if time remained) or move on if the 180-second limit expired.

After completing each trial, participants were asked to fill out the SUS, SIM-TLX, IPQ, NQ. At the end of all three trials, participants completed the SSQ again, followed by an open-ended interview to gain deeper insights into their experiences.

\subsection{Result}
 
\begin{figure}[h]
    \centering
    \includegraphics[width=1\linewidth]{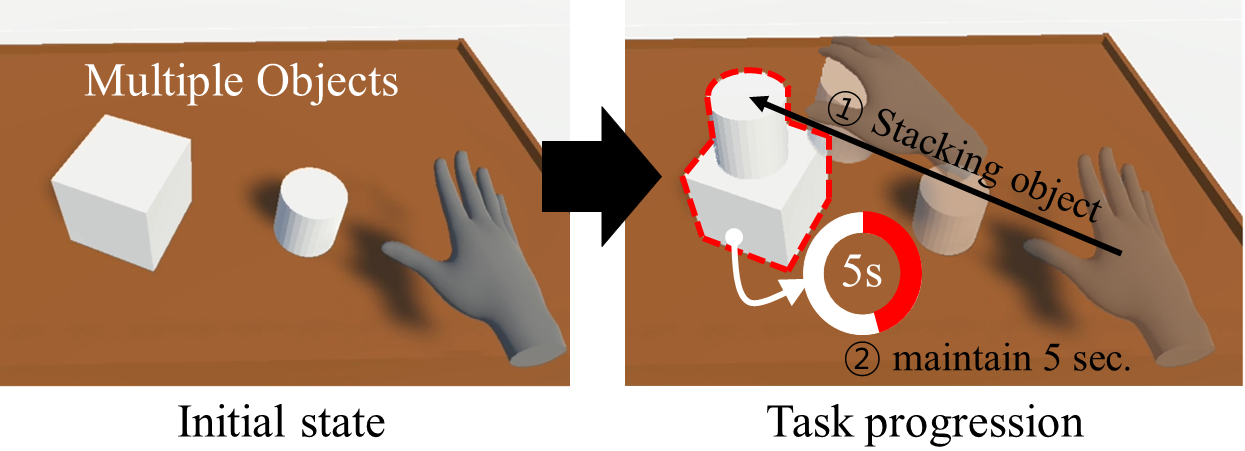}
    \caption{Illustration of the tower-building task: Participants stack objects, and if the tower remains stable for five seconds, the scenario resets with objects repositioned, increasing the number of objects to stack with each success. This process repeats throughout the experiment.}
    \label{fig:Overview_tower_building_task}
\end{figure}

\begin{figure}
    \centering
    \includegraphics[width=1\linewidth]{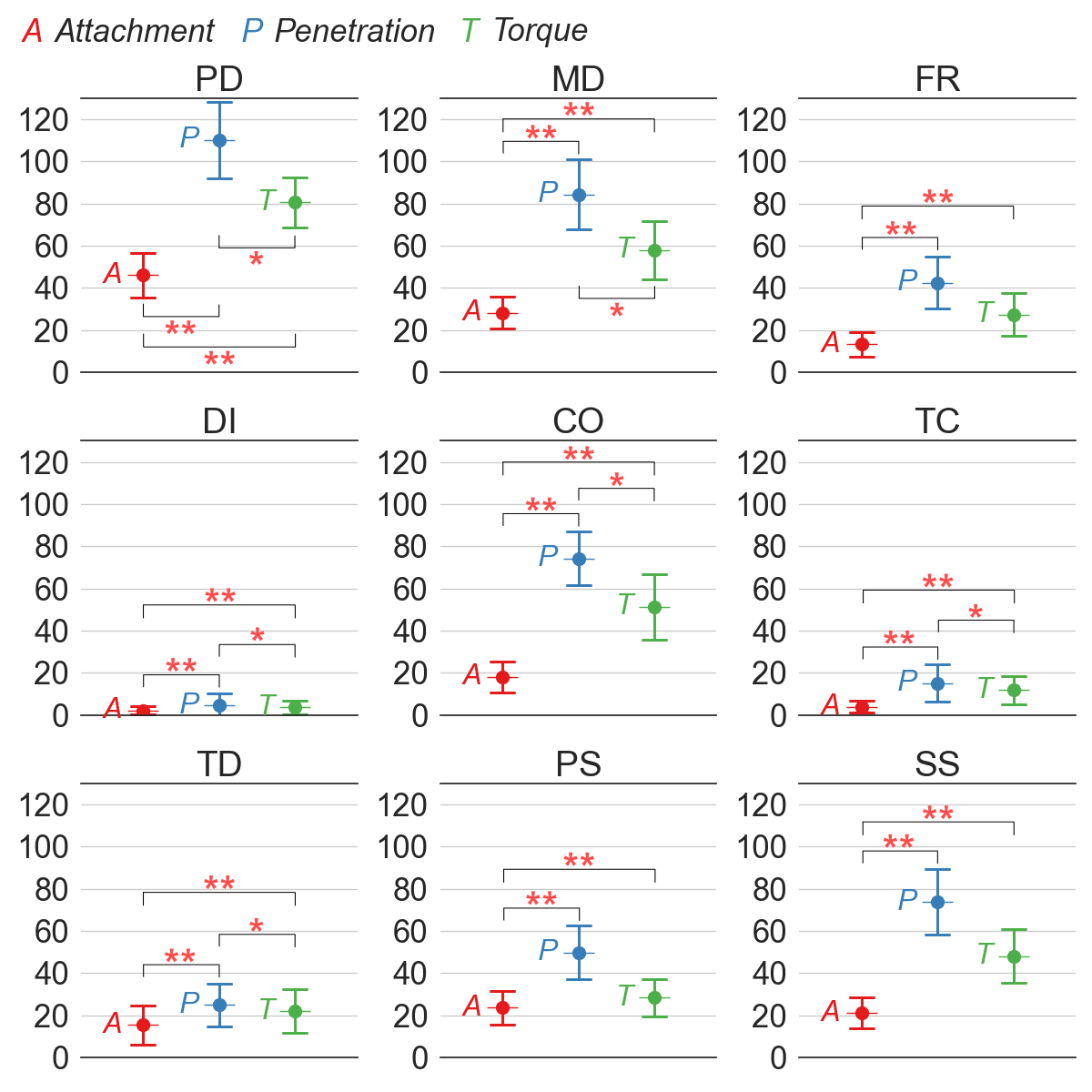}
    \caption{SIM-TLX analysis for the tower-building task.  The graph plots the mean and the standard deviation. Square brackets between groups within the same item indicate the results of the Wilcoxon signed-rank test ($*$ : p $<$ 0.05, $**$ : p $<$ 0.01). }
    \label{fig:SIM-TLX_TowerBuildingTask}
    \vspace{-4mm}
\end{figure}

\begin{table}[h]
    \caption{The recorded performance of participants across the three different test methods in E2.}
    \label{tab:Precise_control_task}
    \centering
    \resizebox{\linewidth}{!}{
    \begin{tabular}{lSSS}
        \Xhline{2\arrayrulewidth}
        {} & {\textit{Attachment}} & {\textit{Penetration}} & \textit{{Torque}} \\
        \Xhline{2\arrayrulewidth}
        Task completion time (s) & 28.594 & 45.449 & 45.483 \\ 
        \hline 
        \# of object drops & 5.908 & 7.136 & 4.247 \\ 
        \hline 
        \# of floors stacked & 6.208 & 2.909 & 4.000 \\ 
        \Xhline{2\arrayrulewidth}
    \end{tabular}
    }
\end{table}

\begin{table}[h]
    \caption{Mean and standard of SIM-TLX in E2.}
    \label{tab:Tower_building_task_TLX}
    \centering
    \resizebox{\linewidth}{!}{%
    \begin{tabular}{l||SSS}
        \Xhline{2\arrayrulewidth}
        {Dimension} & {\textit{Attachment}} & {\textit{Penetration}} & {\textit{Torque}} \\
        \Xhline{2\arrayrulewidth}
        PD & {46.13  $\pm$ 10.55} & {110.25 $\pm$ 18.09} & {80.62 $\pm$ 11.89} \\ 
        \hline 
        MD & {28.34 $\pm$ \phantom{0}7.64} & {\phantom{0}84.34 $\pm$ 16.62} & {58.00 $\pm$ 13.91} \\ 
        \hline 
        FR & {13.33 $\pm$  \phantom{0}5.89} & {\phantom{0}42.50 $\pm$ 12.30} & {27.33 $\pm$ 10.14} \\ 
        \hline 
        DI & {\phantom{0}2.13 $\pm$  \phantom{0}1.90} & {\phantom{00}4.54 $\pm$ \phantom{0}5.46} & {\phantom{0}3.50 $\pm$ \phantom{0}3.06} \\ 
        \hline 
        CO & {18.00 $\pm$  \phantom{0}7.30} & {\phantom{0}74.25 $\pm$ 12.80} & {51.25 $\pm$ 15.46} \\ 
        \hline 
        TC & {\phantom{0}3.83 $\pm$ \phantom{0}2.76} & {\phantom{0}15.00 $\pm$ \phantom{0}8.80} & {11.75 $\pm$ \phantom{0}6.59} \\ 
        \hline 
        TD & {15.50 $\pm$  \phantom{0}9.25} & {\phantom{0}24.88 $\pm$ 10.20} & {21.88 $\pm$ 10.37} \\ 
        \hline 
        PS & {23.54 $\pm$ \phantom{0}7.93} & {\phantom{0}49.79 $\pm$ 12.62} & {28.34 $\pm$ \phantom{0}8.99} \\ 
        \hline 
        SS & {21.00 $\pm$  \phantom{0}7.32} & {\phantom{0}73.79 $\pm$ 15.60} & {48.13 $\pm$ 12.85} \\ 
        \Xhline{2\arrayrulewidth}
    \end{tabular}
    }
\end{table}

\begin{figure*}
    \centering
    \includegraphics[width=1\linewidth]{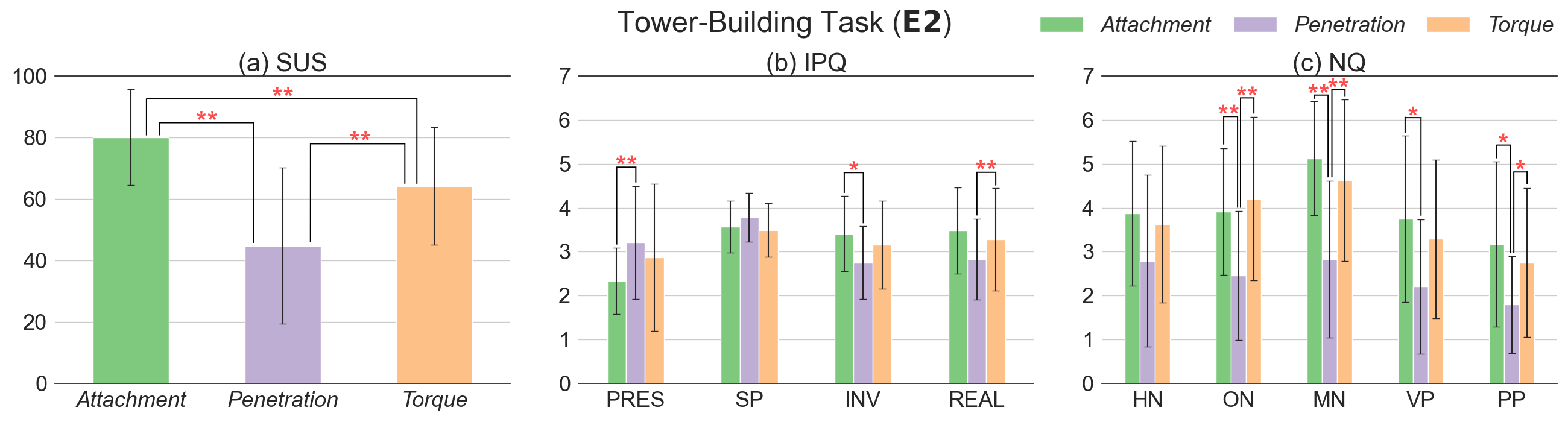}
    \caption{SUS, IPQ, and NQ analysis for the tower-building task. The graph plots the mean and the standard deviation. Square brackets between groups within the same item indicate the results of the Wilcoxon signed-rank test ($*$ : p $<$ 0.05, $**$ : p $<$ 0.01). }
    \label{fig:Subjective_TowerBuildingTask}
    \vspace{-1mm}
\end{figure*}

\subsubsection{Objective Evaluation}
\autoref{tab:Precise_control_task} summarizes the participants' performance recorded across three different test methods in the tower-building task scenario. The performance is evaluated in terms of task completion time, defined as the average time participants took to complete a single stacking, the average number of object drops, and the average number of floors stacked during the trial. 

For the task completion time and number of floors stacked, \textit{Attachment} outperformed both \textit{Penetration} and \textit{Torque}. 
Similar to E1, this superior performance can be attributed to its simplicity, focusing solely on the relative positioning of the hand and the object without requiring users to manage the force applied during grasping.

On the other hand, \textit{Torque} exhibited the best performance regarding the number of object drops, likely due to its ability to enable precise object manipulation at the desired positions with the correct amount of force.
In contrast, \textit{Penetration} resulted in more frequent object drops and a lower number of successfully stacked floors, indicating that accurately picking up and placing objects through mid-air interactions proved to be more challenging for users.


\subsubsection{Subjective Evaluation}
To analyze the collected responses, we first assessed normality using the Shapiro-Wilk test. Since some responses did not follow a normal distribution, we applied the Friedman test to detect statistically significant differences. For post-hoc analysis, we conducted Wilcoxon signed-rank tests.

\autoref{fig:SIM-TLX_TowerBuildingTask} and \autoref{tab:Tower_building_task_TLX} demonstrates SIM-TLX analysis results. Similar to those in E1, \textit{Penetration} consistently scored higher than the other two methods across all dimensions. The results of the Friedman test revealed significant differences across the following dimensions: PD, MD, TD, SS, TC, FR, and PS ($p < 0.01$). Conversely, no significant difference was observed for the DI dimension ($p = 0.155$).
A substantial increase in user workload is also reported when interactions do not involve a controller. Our field observations indicated that users felt the use of \textit{Penetration} more challenging in complex tasks than in simple tasks. This was further supported by the SIM-TLX results, which showed higher scores in E2 compared to E1, reflecting increased workload during the more complex tower-building task. The mid-air hand motion appears to require more nerve to control the desired virtual hand pose and maintain a stable balance.

\autoref{fig:Subjective_TowerBuildingTask}(a) presents the SUS analysis. It revealed a ranking of usability with \textit{Attachment} demonstrating the highest usability ($\mu$ = 80.104, $\sigma$ = $\pm$15.630), followed by \textit{Torque} ($\mu$ = 64.167, $\sigma$ = $\pm$19.149), and \textit{Penetration} ($\mu$ = 44.792, $\sigma$ = $\pm$25.409).
Statistical comparisons revealed that \textit{Attachment} had significantly higher usability scores than both \textit{Penetration} and \textit{Torque}. Additionally, \textit{Torque} showed higher usability scores than \textit{Penetration} ($\chi^2(2) = 29.702$, $p < 0.01$).
As in E1, concentrating solely on the grasping action without accounting for physical controllability, while relying on VR controllers, appears to have a positive impact on usability.

\autoref{fig:Subjective_TowerBuildingTask}(b) reports the IPQ analysis results. Friedman test revealed that there were no significant differences except REAL ($\chi^2(2) = 11.507$, $p < 0.01$). Wilcoxon signed-rank test showed that, \textit{Attachment} ($\mu$ = 3.479, $\sigma$ = $\pm$0.981) and \textit{Torque} ($\mu$ = 3.281, $\sigma$ = $\pm$1.164) was rated higher than \textit{Penetration} ($\mu$ = 2.823, $\sigma$ = $\pm$0.919) in terms of REAL (\textit{Z} = 10.504, \textit{p} $<$ 0.01).

\autoref{fig:Subjective_TowerBuildingTask}(c) presents the naturalness analysis results. 
Generally, users felt unnatural with \textit{Penetration} compared to \textit{Attachment} and \textit{Torque}.
Interestingly, For the ON dimension, \textit{Torque} ($\mu$ = 4.209, $\sigma$ = $\pm$1.865) scored significantly higher than \textit{Attachment} ($\mu$ = 3.917, $\sigma$ = $\pm$1.442) and \textit{Penetration} ($\mu$ = 2.458, $\sigma$ = $\pm$1.474) ($\chi^2(2) = 16.975$, $p < 0.01$).
For the MN dimension, \textit{Attachment} ($\mu$ = 5.125, $\sigma$ = $\pm$1.296) scored higher than \textit{Penetration} ($\mu$ = 2.833, $\sigma$ = $\pm$1.786) and \textit{Torque} ($\mu$ = 4.625, $\sigma$ = $\pm$1.837)($\chi^2(2) = 23.763$, $p < 0.01$). 
For the VP dimension, \textit{Attachment} ($\mu$ = 3.750, $\sigma$ = $\pm$1.894) scored higher than \textit{Penetration} ($\mu$ = 2.208, $\sigma$ = $\pm$1.532) and \textit{Torque} ($\mu$ = 3.292, $\sigma$ = $\pm$1.805) ($\chi^2(2) = 7.461$, $p < 0.05$). 
Lastly, for the PP dimension, \textit{Attachment} ($\mu$ = 3.167, $\sigma$ = $\pm$1.880) scored higher than \textit{Penetration} ($\mu$ = 1.792, $\sigma$ = $\pm$1.103) and \textit{Torque} ($\mu$ = 2.750, $\sigma$ = $\pm$1.700) ($\chi^2(2) = 11.057$, $p < 0.01$).
Unlike in E1, no significant differences were observed in the HN dimension.


There was no statistically significant change in SSQ scores before and after the experiment.

\section{Discussion}
This section discusses the advantages and disadvantages of each VR hand interaction method based on evaluation results and field observations. From these discussions, we derive design guidelines and suggest directions for future improvement. Finally, we address our two research questions.

\subsection{Discussion about \textit{Attachment}}
The \textit{Attachment} generally outperformed the other methods both in objective and subjective evaluations.
Despite providing the lowest level of visual and physical plausibility, this approach proved highly effective and intuitive for hand manipulation tasks. 

One key advantage is its limited controllability, which enables users to complete tasks quickly and effortlessly. Several participants noted, ``\textit{Being able to easily grab and lift the object, and then release it whenever I wanted, made it realistic to complete the task}” (P7, P12, P13, P15).

Another strength of the \textit{Attachment} is its smooth finger movements and stable contact with virtual objects. As some participants observed, ``\textit{It felt natural for my fingers to bend smoothly and make stable contact with the object when I pressed a button to grasp it}” (P1, P9).

However, many participants reported difficulties and unnatural experiences when interacting with small or complex-shaped objects. One user stated, ``\textit{It was hard to lift small or complex-shaped objects, and getting my palm very close to grab them felt awkward and inconvenient}” (P1, P2, P4, P6, P8, P12, P17, P20). A potential solution suggested by a participant involves allowing finger bending even when no object is detected: ``\textit{If we could bend our fingers even when no object is detected and grasp objects just with our fingers, such issue will be mitigated. Right now, this cannot be used for grabbing various objects}” (P2).

Participants also noted that the hand shapes appeared somewhat unnatural. One user mentioned, ``\textit{The hand shape looked like it couldn't actually lift anything when the object has complex shapes, so it seemed weird when the object was lifted}” (P4, P14, P23, P24). 
Similarly, some found it appeared unnatural to adjust all five fingers simultaneously using the controller: ``\textit{Grabbing the object with all five fingers moving together felt like the virtual hand was more like a robot's than my own}” (P5, P21).

Interestingly, the experience of attaching objects to the hand after establishing contact points received mixed feedback. Some participants felt it was unnatural: ``\textit{Sometimes, it felt like the object was just stuck to the hand and moved with it, which seemed odd}” (P3, P21). Others, however, saw a practical benefit: ``\textit{It may not look natural, but this easier way of grabbing objects will probably be used more to improve accessibility and usability in VR}” (P14).

In summary, future research and commercial implementations should retain the simplicity and acceptable realism of the \textit{Attachment}, while addressing its limitations in handling complex object scenarios and reducing the unnatural sensations caused by awkward hand shapes and finger movements.

\subsection{Discussion about \textit{Penetration}}
\textit{Penetration} consistently received lower scores across most evaluations, with the exception of sense of presence. Lower success rates, longer task completion times, and poor SIM-TLX scores indicated that this method is less efficient for hand-object manipulation tasks. 

The primary issue lies in the lack of stability when lifting objects. As five participants noted, ``\textit{After grabbing the object, it sometimes vibrated in my hand, making it hard to move in the direction I wanted}" (P1, P2, P6, P9, P18). This instability arises because the method generates contact points with each finger when grasping, but does not account for the equilibrium of forces needed to maintain stable pressure. For example, typically, the thumb, index, and middle fingers exert the most force, while the ring and pinky fingers contribute less. When moving an object, the applied force also changes. However, \textit{Penetration} does not account for dynamic uneven force distribution across fingers.

This problem is particularly pronounced when interacting with very small or large objects. One participant remarked, ``\textit{Lifting small or large objects was challenging and frustrating}” (P10, P20, P21). 
The difficulty arises from the recent study's naive design, which converts the degree of penetration into applied force based solely on distance.
 
Another drawback is the fatigue caused by mid-air hand movements and the effort needed to curl fingers when grasping virtual objects. One participant noted, ``\textit{After trying to lift the object several times, my shoulder started hurting from all the effort}” (P12). Another added, ``\textit{Maintaining a stable grasp with mid-air hand movements tires me out quickly}” (P15).

Although the \textit{Penetration} was designed to provide the highest level of visual realism by accurately replicating real-world hand movements, it unexpectedly received low scores for perceived naturalness.
This discrepancy likely arises from differences in interaction ways that deviate from real-world cases.
Participants expressed this sentiment: ``\textit{Unlike real life, where grabbing objects is easy, curling my fingers in VR was difficult, and felt disconnected}” (P2, P4, P12, P15).

Furthermore, the replication is limited by the method's reliance on converting the degree of object penetration into force. This approach does not accommodate interactions where there is no space to penetrate or where penetration is unnecessary, limiting users from executing all desired interactions. This restriction has been pointed out as a significant drawback.

On a positive note, the method received favorable feedback for its ability to closely mimic real-life hand tracking. As one participant stated, ``\textit{It was great that the virtual hand tracked my fingers accurately, making the grasp feel natural}” (P9, P12, P17). One participant suggested that if the stability issues could be resolved, the method would be the most promising for hand-object manipulation: ``\textit{If objects can be grasped effectively and stably, hand tracking seems like the best approach}” (P22).

In conclusion, future research and commercial implementations should retain the realism and fine finger control provided by accurate hand tracking. Specifically, while the current approach relies on the contact point at the moment of touch for visual mapping, improving visual plausibility will necessitate refining the visual mapping to depict hand poses to naturally conform to the grasped object. Furthermore, the inaccuracies caused by the hand tracking jitters can be mitigated by employing denoising technique~\cite{luo2024physics}.

From the perspective of physical plausibility, the consideration of force equilibrium is paramount. To ensure stability and prevent object wobbling during grasping, the forces applied must be balanced around the object’s center of mass, with equal and opposing forces. Recent advancements, such as the grasping determination phase demonstrated by Wang et al. \cite{wang2024realistic}, offer promising methodologies for maintaining force equilibrium under similar conditions.

\subsection{Discussion about \textit{Torque}}
The \textit{Torque} occupies an intermediate position between the other two methods. It demonstrated a level of naturalness and workload comparable to that of the \textit{Attachment}.

One of the key advantages of the \textit{Torque} is its ability to effectively grasp complex-shaped objects, such as allowing users to hook their fingers around the handle of a teapot-shaped object. As one participant remarked, ``\textit{It was surprising to be able to hook my fingers onto the teapot handle}” (P3).

However, the method still lagged behind \textit{Attachment} in terms of task completion time and success rate while its usability was rated as moderate. 
Some participants noted that it did not provide sufficient support for very small or large objects. One participant stated, ``\textit{I felt like I couldn’t apply enough force when lifting small objects}” (P8). 
Additionally, the method struggled to show realistic finger movements for certain objects, particularly with the index and middle fingers. 
As one participant noted, ``\textit{Virtual hands couldn’t use index and middle fingers properly, which made it unnatural when grasping some objects}” (P14). 
Another added, ``\textit{When I pressed the button to grab the object, the fingers moved too slowly, which was frustrating and felt unnatural}” (P6).
These limitations likely stem from the model’s inability to consistently learn across a wide variety of object shapes and sizes.

Regarding force control, participant feedback was mixed. Some felt a strong sense of force applied to the object, contributing to their high ratings for naturalness. Several participants noted that the realistic sensation of pressing the controller’s trigger button contributed to realism: ``\textit{Pressing the button firmly gave me a realistic sense of gripping the object}” (P8, P10, P21).

However, several participants reported experiencing unnatural sensations of slipping, especially when interacting with very small or large objects. One participant mentioned, ``\textit{It felt unnatural because the object kept slipping and wouldn’t lift}” (P10).

In conclusion, future research and commercial implementations should retain the \textit{Torque}’s versatility in grasping complex-shaped objects while addressing its limitations in force control and stability. From the perspective of visual plausibility, it is crucial to properly train finger bending angles during object grasping, particularly when buttons are pressed, to prevent specific fingers from inadvertently pushing the object away. A promising strategy to achieve this involves implementing real-time inverse kinematics adjustments based on contact dynamics. Such adjustments can also mitigate discrepancies caused by tracking inaccuracies or unexpected object responses.

Regarding physical plausibility, a key challenge identified in user interviews is the underutilization of the index and middle fingers, which compromises overall grasp stability. To address this issue, weighted priorities can be applied to the joints of each finger during training, ensuring that finger movements contribute meaningfully to stable and effective grasps.

\section{Conclusion}
In this paper, we explore three recent VR hand interaction methods—\textit{Attachment}, \textit{Penetration}, and \textit{Torque}—analyzing their strengths and weaknesses across various scenarios. Our goal is to identify the technical features that influence their performance and to propose development guidelines for enhancing user experience. To achieve this, we implemented these methods and evaluated them through two user studies.

The results highlight that \textit{Attachment} is the most suitable for general commercial use due to its simplicity, predictable behavior, and ease of implementation. While it sacrifices physical and visual realism, its straightforward control delivers a consistently positive user experience. In contrast, \textit{Penetration} and \textit{Torque} exhibit limitations in physical controllability and visual realism, making them less viable for immediate deployment. Users reported issues such as instability, fatigue, and unnatural interactions when handling objects of various shapes.

Despite these challenges, \textit{Penetration} and \textit{Torque} exhibit significant potential for next-generation VR interactions. With further refinement, \textit{Penetration} could offer highly versatile hand interactions by closely mirroring real hand movements. Key improvements include optimizing force distribution, mitigating fatigue during mid-air interactions, and expanding support for non-penetrative interactions. Similarly, \textit{Torque} holds promise for boader adaptability across diverse VR scenarios if it achieves visually realistic hand animations and physically plausible force dynamics. Its reliance on the controller enhances accessibility, and advancements in AI-driven control system may soon address its current limitations, paving the way for more seamless and immersive VR experiences.

However, this study has certain limitations. It was conducted exclusively using the Oculus Quest 2, which may restrict the applicability of the findings to other VR systems. The participant sample also demonstrated gender and low age biases, and the sample size ($N$ = 24) may have been insufficient for robust statistical analysis. Furthermore, the study's scope was limited to single-object interaction tasks. Future research should explore diverse hardware platforms, incorporate more diverse participant demographics, and examine multi-object interaction scenarios to develop a more comprehensive understanding of VR hand interactions.

We believe this study provides valuable insights into the current state and future potential of VR hand interaction methods, offering a foundation for developing more realistic and advanced interaction technologies.

\acknowledgments{
This work was supported by ICT Creative Consilience Program through the Institute of Information \& Communications Technology Planning \& Evaluation(IITP) grant funded by the Korea government(MSIT)(IITP-2025-RS-2020-II201819) and Institute of Information \& communications Technology Planning\& Evaluation (IITP) grant funded by the Korea government(MSIT) (No.RS-2020-II200861).}

\bibliographystyle{abbrv-doi} 

\end{document}